\documentclass[12pt]{iopart}
\usepackage{iopams}  
\usepackage{epsfig}

\usepackage{bbm}
\usepackage{color}
\usepackage{ulem}

\usepackage{enumitem}
\usepackage{graphicx}
\usepackage{subfigure}
\usepackage{graphics}

\newcommand{\be}{\begin{equation}}
\newcommand{\ee}{\end{equation}}
\newcommand{\bea}{\begin{eqnarray}}
\newcommand{\eea}{\end{eqnarray}}

\def\({\left(}
\def\){\right)}

\newcommand{\bmat}{\xi}  

\newcommand{\meff}{m_\phi}       
\newcommand{\mmax}{m_\mathrm{max}}
\newcommand{\Mpl}{M_\mathrm{Pl}} 
\newcommand{\phibulk}{\phi_\mathrm{m}}   
\newcommand{\rholab}{\rho_\mathrm{lab}}
\newcommand{\rhomat}{\rho} 
\newcommand{\Vloop}{V_\mathrm{1-loop}}

\def\lsim{ \lower .75ex \hbox{$\sim$} \llap{\raise .27ex \hbox{$<$}} }
\def\gsim{ \lower .75ex \hbox{$\sim$} \llap{\raise .27ex \hbox{$>$}} }

\begin{document}

\title{Chameleon Field Theories}

\author{Justin Khoury}

\address{Center for Particle Cosmology, Department of Physics and Astronomy, University of Pennsylvania,
Philadelphia, Pennsylvania 19104, USA}
\ead{jkhoury@sas.upenn.edu}
\begin{abstract}
Chameleons are light scalar fields with remarkable properties. Through the interplay of self-interactions and coupling to matter, chameleon particles have a mass that depends
on the ambient matter density. The manifestation of the fifth force mediated by chameleons therefore depends sensitively on their environment, which makes for a rich phenomenology. In this article,
we review two recent results on chameleon phenomenology. The first result a pair of no-go theorems limiting the cosmological impact of chameleons and their generalizations:
i) the range of the chameleon force at cosmological density today can be at most $\sim$~Mpc; ii) the conformal factor relating Einstein- and Jordan-frame scale factors is
essentially constant over the last Hubble time. These theorems imply that chameleons have negligible effect on the linear growth of structure, and cannot account for the
observed cosmic acceleration except as some form of dark energy. The second result pertains to the quantum stability of chameleon theories. We show how requiring 
that quantum corrections be small, so as to allow reliable predictions of fifth forces, leads to an upper bound of $m < 0.0073 (\rhomat / 10~{\rm g\, cm}^{-3})^{1/3}$eV for gravitational strength coupling, whereas fifth force experiments place a lower bound of $m>0.0042$\,eV. An improvement of less than a factor of two in the range of fifth force experiments could test all classical chameleon field theories whose quantum corrections 
are well-controlled and couple to matter with nearly gravitational strength regardless of the specific form of the chameleon potential.  
\end{abstract}

\maketitle

\section{Introduction}

The $\Lambda$-Cold Dark Matter ($\Lambda$CDM) model, which features a cosmological constant as dark energy and weakly-interacting particles as dark matter,
has emerged as the standard model of cosmology~\cite{Jain:2010ka}. The empirical success of the model lies in its parsimony --- only a handful of parameters are required
to fit all known cosmological observations. It is therefore highly predictive: the $\Lambda$CDM expansion and growth histories are tightly correlated, leaving
essentially no wiggle room to account for possible discrepancies.

In the coming years, the $\Lambda$CDM model will be confronted by a host of increasingly powerful probes of the large scale structure. Experiments
like the Dark Energy Survey~\cite{DES}, BigBOSS~\cite{BigBOSS}, the Large Synoptic Survey Telescope~\cite{LSST} and EUCLID~\cite{EUCLID} will test the standard model predictions 
with unprecedented accuracy. These precision-test experiments may well reveal the existence of ``beyond-the-standard-model" physics, in the form of new
light degrees of freedom in the dark sector.

Whatever the nature of these new degrees of freedom, they must somehow effectively decouple from matter on solar system/laboratory scales, to conform with
the stringent constraints from local tests of gravity and reproduce General Relativistic predictions to sufficient accuracy~\cite{Will:2005va}. This can be achieved through {\it screening mechanisms},
which rely on the high density of the local environment (relative to the mean cosmological density) to suppress deviations from standard gravity.
The manifestation of the new degrees of freedom (generally scalar fields) therefore depends sensitively on their environment, which in turn leads to striking experimental signatures. 

To date, only three successful screening mechanisms have been proposed:

\begin{itemize}

\item {\it The Chameleon Mechanism}~\cite{Khoury:2003aq,Khoury:2003rn,Gubser:2004uf,Brax:2004qh,Mota:2006ed,Mota:2006fz} operates whenever a scalar field couples to matter in such a way that its effective mass depends on the local matter density. Deep in space, where the mass density is low, the scalar is light and mediates a fifth force of gravitational strength, but near the Earth, where experiments are performed, and where the local density is high, it acquires a large mass, making its effects short range and hence unobservable. 

\item {\it The Vainshtein Mechanism}~\cite{Vainshtein:1972sx,ArkaniHamed:2002sp,Deffayet:2001uk} relies on derivative couplings of a scalar field becoming large in the vicinity of massive sources. These non-linearities crank up the kinetic term of perturbations, thereby weakening their interactions with matter. This mechanism is essential to the viability of massive gravity~\cite{deRham:2010ik,deRham:2010kj,Hinterbichler:2011tt} and its extensions~\cite{D'Amico:2012zv,Gabadadze:2012tr}, the Dvali-Gabadadze-Porrati model~\cite{Dvali:2000hr,Dvali:2000xg}, cascading gravity~\cite{deRham:2007xp,deRham:2007rw,deRham:2009wb,deRham:2010rw,Agarwal:2009gy,Agarwal:2011mg}, and galileon theories~\cite{Luty:2003vm,Nicolis:2008in,Deffayet:2009wt,Deffayet:2009mn,deRham:2010eu,Padilla:2010de,Hinterbichler:2010xn,Chow:2009fm,Silva:2009km,Goon:2011qf,Goon:2011uw}. 

\item {\it The Symmetron Mechanism}~\cite{Hinterbichler:2010es}, based on earlier work by~\cite{Olive:2007aj,Pietroni:2005pv}, relies on the vacuum expectation value (VEV) of a scalar field that depends on the local mass density, becoming large in regions of low mass density, and small in regions of high mass density. The coupling of the scalar to matter is proportional to the VEV, so that the scalar couples with gravitational strength in regions of low density, but is decoupled and screened in regions of high density. The cosmology of symmetrons was studied in~\cite{Hinterbichler:2011ca,Brax:2011pk}. A closely related mechanism is the varying-dilaton mechanism~\cite{Brax:2011ja}.

\end{itemize}

Independent of cosmology, screening mechanisms are motivated by the vast experimental effort aimed at testing the fundamental nature of gravity on a wide range of scales, from laboratory to solar system to extra-galactic scales. Viable screening theories make novel predictions for local gravitational experiments. The subtle nature of these signals have forced experimentalists to rethink the implications of their data and have inspired the design of novel experimental tests. The theories of interest thus offer a rich spectrum of testable predictions for ongoing and near-future tests of gravity.

The idea that the manifestation of a fifth force is sensitive to the environment has spurred a lot of activity. Astrophysically, chameleon scalars affect the
internal dynamics~\cite{Hui:2009kc,Jain:2011ji} and stellar evolution~\cite{Chang:2010xh,Davis:2011qf,Jain:2012tn} in dwarf galaxies in void or
mildly overdense regions. In the laboratory, chameleons have motivated multiple experimental efforts aimed at searching for chameleon signatures:

\vspace{0.1cm}
\noindent $\bullet$ The E\"ot-Wash experiment searches for deviations from the inverse-square-law at distances $\mathrel{\mathstrut\smash{\ooalign{\raise2.5pt\hbox{$>$}\cr\lower2.5pt\hbox{$\sim$}}}} 50\;\mu$m. Based on theoretical predictions~\cite{Upadhye:2006vi}, the E\"ot-Wash group analyzed their data to constrain part of the chameleon parameter space~\cite{Adelberger:2006dh}.

\vspace{0.1cm}
\noindent $\bullet$ If chameleons interact with the electromagnetic field via $e^{\beta_\gamma\phi}F_{\mu\nu}F^{\mu\nu}$, then photons traveling in a magnetic field will undergo photon/chameleon oscillations. The CHameleon Afterglow SEarch (CHASE) experiment~\cite{Chou:2008gr,Steffen:2010ze} has looked for an afterglow from trapped chameleons converting into photons. Similarly, the  Axion Dark Matter eXperiment (ADMX) resonant microwave cavity was used recently to search for chameleons~\cite{Rybka:2010ah}. Photon-chameleon mixing can occur deep inside the Sun~\cite{Brax:2010xq} and affect the spectrum of distant astrophysical objects~\cite{Burrage:2008ii}. 

Through a nice analogy between chameleon screening and electrostatics~\cite{JonesSmith:2011tn,Pourhasan:2011sm}, a novel experimental test of chameleons has been proposed
recently that would exploit an enhancement of the scalar field near the tip of pointy objects (a``lightning rod" effect)~\cite{JonesSmith:2011tn}. See~\cite{Brax:2009ey} for a
discussion of collider signatures, and~\cite{Upadhye:2012rc} for signatures of symmetron in the laboratory.

The most striking signature of chameleons can be found by testing gravity in space. Because the screening condition depends on the ambient density, small bodies that are screened in the
laboratory may be unscreened in space. This leads to striking predictions for future satellite tests of gravity, such as the planned MicroSCOPE mission~\cite{MICROSCOPE} and STE-QUEST~\cite{STE}. In particular,
chameleons can result in violations of the (weak) Equivalence Principle in orbit with $\eta \equiv \Delta a/a \gg 10^{-13}$, in blatant conflict with laboratory constraints. Similarly,
the total force --- gravitational + chameleon-mediated --- between unscreened particles can be ${\cal O}(1)$ larger than in standard gravity, which would appear as ${\cal O}(1)$ deviations
from the value of $G_{\rm N}$ measured on Earth.

Chameleons and symmetrons were constructed from a bottom-up approach, with the potential and matter coupling rigged up to hide the scalar locally.
To put these ideas on firmer theoretical footing, ideally one would like to see an explicit UV-complete realization in string theory. This is in principle possible,
since chameleons and symmetrons are described by canonical scalar fields with self-interaction potentials, and hence are compatible with a Wilsonian UV completion.\footnote{This is in contrast with galileons.
Since galileons generally propagate superluminally on certain backgrounds, their UV completion cannot be a standard local quantum field theory or perturbative string theory~\cite{Adams:2006sv}. See~\cite{Khoury:2010gb,Khoury:2011da,Koehn:2012ar,Koehn:2012te,Koehn:2013hk} for recent embeddings of galileons and general higher-derivative scalar theories in supersymmetry and supergravity.}
As a first step in this direction,~\cite{Hinterbichler:2010wu} presented a scenario for embedding the chameleon screening mechanism within supergravity/string theory compactifications. (See~\cite{Nastase:2013ik,Nastase:2013los} for related work, and~\cite{Brax:2011qs,Brax:2012mq} for other supersymmetric extensions of chameleon/symmetron/dilaton screening.) In this approach, the chameleon scalar field is identified with a certain function of the volume modulus of the extra dimensions.  In follow-up work,~\cite{Hinterbichler:2013we} extended the scenario and showed, with suitable generalization of the superpotential and Kahler potential, that the volume modulus can also drive slow-roll inflation in the early universe.

In this article, we focus primarily on chameleon field theories. Specifically, after reviewing the basics of the chameleon mechanism in Sec.~\ref{reviewsec}, we turn our attention to two recent developments of interest, detailed in~\cite{Wang:2012kj} and~\cite{Upadhye:2012vh}, respectively. We first discuss in Sec.~\ref{chamnogo} a pair of important no-go theorems limiting its cosmological impact: i) the Compton wavelength of the chameleon can be at most Mpc at present cosmic density, which restricts its impact to non-linear scales;  ii) the conformal factor relating Einstein- and Jordan-frame scale factors is essentially constant over the last Hubble time, which precludes the possibility of self-acceleration. We then estimate in Sec.~\ref{chamquantum} quantum corrections to the chameleon potential. Focusing on scalar loops, we show that requiring quantum corrections to be under control, so as to allow reliable predictions for fifth force experiments, places an upper bound of $m < 0.0073 (\rhomat / 10~{\rm g\, cm}^{-3})^{1/3}$eV for gravitational strength coupling whereas fifth force experiments place a lower bound of $m>0.0042$\,eV. An improvement of less than a factor of two in the range of fifth force experiments could test all classical chameleon field theories whose quantum corrections are under control and couple to matter with nearly gravitational strength regardless of the specific form of the chameleon potential.  


\section{A Brief Review of Chameleons}
\label{reviewsec}

Chameleon scalar fields mediate a fifth force of gravitational strength between matter particles, with a range that decreases with increasing ambient matter density,
thereby avoiding detection in regions of high density~\cite{Khoury:2003aq,Khoury:2003rn,Gubser:2004uf,Brax:2004qh}. This is achieved within the framework of a general scalar-tensor theory
in the Einstein frame, with scalar potential $V(\phi)$ and matter coupling $A(\phi)$:\footnote{One can in fact allow different couplings to the various matter fields, thereby explicitly violating the Equivalence Principle. For the purpose of this article, we focus on the simplest case of a universal, conformal coupling.}
\be
S = \int {\rm d}^4x \sqrt{-g} \left(\frac{R}{16\pi G_{\rm N}} - \frac{1}{2}(\partial\phi)^2 - V(\phi)\right) + S_{\rm m}[g^{\rm J}]\,,
\label{Scham}
\ee
where 
\be
g_{\mu\nu}^{\rm J} = A^2(\phi)g_{\mu\nu}
\label{confrel}
\ee
is the Jordan-frame metric. Matter fields described by $S_{\rm m}$ couple to $\phi$ through the conformal factor $A(\phi)$ implicit in $g_{\mu\nu}^{\rm
  J}$. The acceleration of a {\it test particle} is influenced by the scalar:
\be
\vec{a} = -\vec{\nabla}\Phi_{\rm N} - \frac{{\rm d}\ln A(\phi)}{{\rm d}\phi} \vec{\nabla}\phi = -\vec{\nabla}\bigg(\Phi_{\rm N} + \ln A(\phi)\bigg)\,,
\label{acc}
\ee
where $\Phi_{\rm N}$ is the (Einstein-frame) Newtonian potential. The equation of motion for $\phi$ that derives from this action is
\be
\Box \phi = V_{,\phi} + A_{,\phi} \, \rho\,,
\label{phigen}
\ee
where the matter is assumed to be non-relativistic,
and $\rho$ is related to the Einstein- and Jordan-frame
matter densities by $\rho = \rho_{\rm E} / A = A^3 \rho_{\rm J}$ ---
defined such that $\rho$ is conserved in the usual sense in Einstein frame. An alternative form of the $\phi$ equation of motion is
useful for comparing against the Poisson equation for $\Phi_{\rm N}$:
\be
\Box \varphi = 8 \pi G_N (V_{,\varphi} + \alpha A \rho) \, \, ; \qquad
\alpha \equiv {{\rm d} {\,\rm ln\,} A \over {\rm d}\varphi} = M_{\rm Pl}  {{\rm d}
  {\,\rm ln\,} A \over {\rm d} \phi}  \,,
\label{alphadef}
\ee
where $\varphi \equiv \phi/M_{\rm Pl}$, $M_{\rm Pl} \equiv (8\pi
G_{\rm N})^{-1/2}$, and $\alpha$ quantifies the dimensionless
scalar-matter coupling, with $\alpha \sim {\cal O}(1)$ meaning gravitational strength.

Thus, because of its coupling to matter fields, the scalar field is affected by the ambient matter density, and is governed by an effective potential
\be
V_{\rm eff}(\phi) = V(\phi) +A(\phi)\rho \,.
\label{Veffcham}
\ee
For suitably chosen $V(\phi)$ and $A(\phi)$, this effective potential can develop a minimum  at some finite field value $\phi_{\rm min}$ in the presence of background matter density.
For simplicity, we assume that this minimum is unique, within the field range of interest. Further, it is assumed $\phi_{\rm min}$ varies monotonically
with $\rho$, say, ${\rm d}\phi_{\rm min}/{\rm d}\rho \le 0$ --- this is useful for implementing the idea that properties of the
scalar field vary systematically with the ambient density.  Differentiating the equilibrium condition $V_{, \phi} + A_{, \phi} \rho = 0$ with respect to $\phi_{\rm min}$, it is straightforward to
show that 
\be
\frac{{\rm d}\phi_{\rm min}}{{\rm d}\rho} = - \frac{A_{,\phi}(\phi_{\rm min})}{m^2_\phi}\,,
\label{dphimin1}
\ee
where 
\be
m^2_\phi \equiv V_{\rm eff} {}_{,\phi\phi} (\phi_{\rm min}) =
V_{,\phi\phi} (\phi_{\rm min}) + A_{,\phi\phi} (\phi_{\rm min}) \rho
\label{mdef}
\ee
is assumed
non-negative for stability.
This means $A$ must be monotonically increasing --- hence
$V$ must be monotonically decreasing --- with $\phi$, at least over
the field range of interest.
A corollary is that $V(\phi_{\rm min}(\rho))$ and $A(\phi_{\rm min}(\rho))$ are respectively monotonically increasing and decreasing
functions of $\rho$.

A prototypical chameleon potential satisfying these properties is the inverse power-law form, $V(\phi) = M^{4+n}/\phi^n$.\footnote{Potentials with {\it positive} powers of the field, $V(\phi) \sim \phi^{2s}$ with $s$ an integer $\geq 2$, are also good candidates for chameleon theories~\cite{Gubser:2004uf}.} For the coupling function, a generic form that makes contact with Brans-Dicke theories is $A(\phi) \approx 1 + \xi\phi/M_{\rm Pl}$,
where we have used the fact that $\phi\ll M_{\rm Pl}$ over the relevant field range. The constant parameter $\xi$ is implicitly assumed to be~${\cal O}(1)$, corresponding to gravitational strength coupling. 
The effective potential in this case is therefore given by, up to an irrelevant constant,
\be
V_{\rm eff}(\phi) = \frac{M^{4+n}}{\phi^n} + \xi \frac{\phi}{M_{\rm Pl}} \rho\,.
\ee
For $\xi> 0$, this displays a minimum at $\phi_{\rm min} \sim \rho^{-1/(n+1)}$. It follows that the mass of small fluctuations around the minimum, $m^2_{\phi}\sim \rho^{(n+2)/(n+1)}$, is an increasing function of the background density, as desired. The tightest constraint on the model comes from laboratory tests of the inverse square law, which set an upper limit of $\approx 50\;\mu$m
on the fifth-force range assuming gravitational strength coupling~\cite{Adelberger:2006dh}. Modeling the chameleon profile in the E$\ddot{{\rm o}}$t-Wash set-up, and taking into account
that torsion-balance measurements are performed in vacuum, this constraint translates to an upper bound on $M$~\cite{Khoury:2003aq,Khoury:2003rn}
\begin{equation}
M\; \lower .75ex \hbox{$\sim$} \llap{\raise .27ex \hbox{$<$}}\; 10^{-3}\;\;{\rm eV}\,,
\label{Mlim}
\end{equation}
which, remarkably, coincides with the dark energy scale. This also ensures consistency with all known constraints on deviations from General Relativity, including post-Newtonian tests in the solar system and binary pulsar observations~\cite{Khoury:2003aq,Khoury:2003rn}. 

The density-dependent mass immediately results in a further decoupling effect outside sufficiently massive objects, due to the so-called {\it thin-shell} effect. 
This effect can be understood intuitively as follows. If the object is sufficiently massive such that deep inside the object the chameleon minimizes 
the effective potential for the interior density, then the mass of chameleon fluctuations is relatively large inside the object. As a result, the contribution from the core to the exterior
profile is Yukawa-suppressed. Only the contribution from within a thin shell beneath the surface contributes significantly to the exterior profile. In other words, since the chameleon effectively couples
only to the shell, whereas gravity of course couples to the entire bulk of the object, the chameleon force on an exterior test mass is suppressed compared to the gravitational force.
See~\cite{JonesSmith:2011tn,Pourhasan:2011sm} for a nice electrostatic analogy of the thin-shell effect.

An object is therefore said to be screened if the scalar force it sources is everywhere suppressed relative to the gravitational force.
Assuming spherical symmetry, for simplicity, according to (\ref{acc}) this means that
\be
\frac{{\rm d} \ln A(\phi)}{{\rm d}r} \lesssim \frac{{\rm d}\Phi_{\rm N}}{{\rm d}r}\,.
\label{screenedforce}
\ee
Integrating from inside to outside the object, we have
\be
{\,\rm ln \,} \left[ 
{A(\phi_{\rm out}) \over A(\phi_{\rm in})} \right]
\lesssim \Delta \Phi_{\rm N}\,.
\ee
Here, `inside' means the origin $r=0$; `outside' means sufficiently far out such that
$\phi_{\rm out}$ is the equilibrium value for the ambient density. 

To satisfy solar system tests, we typically demand that the Milky Way galaxy is screened.
The gravitational potential is $\Phi_{\rm N} \sim -10^{-6}$, and the ambient field value in this case
is set by today's cosmic mean density: $\phi_{\rm out} = \phi_{z = 0}$. Thus the screening condition is
\be
{\,\rm ln\,} \left[ {A(\phi_{z=0}) \over A(\phi_{\rm in-MW})} \right]
\lesssim 10^{-6} \, .
\label{MWscreen}
\ee
This inequality will be key in proving the no-go theorems of Sec.~\ref{chamnogo}.
It makes clear that it is the gravitational potential of the object in
question, as opposed to its density alone, that ultimately determines whether it is
screened or not. 


\section{Chameleon No-Go Theorems}
\label{chamnogo}

In this Section we review two theorems limiting the extent to which chameleons can impact cosmological observations~\cite{Wang:2012kj}.
The theorems apply to a broad class of chameleon, symmetron and varying-dilaton theories.

The first theorem pertains to the possibility of {\it self-acceleration}. The general action~(\ref{Scham}) involves two metrics: the Jordan-frame metric $g_{\mu\nu}^{\rm J}$, to which matter fields couple minimally, and the Einstein-frame metric $g_{\mu\nu}$, which is by definition governed by the Einstein-Hilbert action. Although the two frames are physically equivalent, cosmological observations are implicitly performed in Jordan frame,
where the masses of particles are constant. Meanwhile, the statement that we need some form of dark energy to drive cosmic acceleration is an Einstein frame statement, where
the Friedmann equation takes its standard form. 

By self-acceleration, we mean accelerated expansion in the Jordan frame, while the Einstein-frame expansion rate is {\it not} accelerating.
This is a sensible definition, for the lack of acceleration in Einstein frame is equivalent to the lack of dark energy. In self-accelerating theories, the observed (Jordan-frame) cosmic acceleration stems entirely
from the conformal transformation~(\ref{confrel}), {\it i.e.}, a genuine modified gravity effect.  Clearly a necessary condition for self-acceleration is that the conformal factor $A(\phi)$ varies by at least ${\cal O}(1)$ over the last Hubble time. We instead find for chameleon-like theories
\be
\frac{\Delta A}{A} \ll 1\,,
\label{Abound}
\ee
ruling out the possibility of self-acceleration. Jordan- and Einstein-frame metrics are indistinguishable, and cosmic acceleration
requires a negative-pressure component. 

The second theorem is an upper bound on the chameleon Compton wavelength at present cosmological density:\footnote{A proof of this result also appeared independently in~\cite{Brax:2011aw}, following a different approach.}
\be
m_\phi^{-1}(\phi_0) \lesssim {\rm Mpc}\,.
\label{mbound}
\ee
Since the chameleon force is Yukawa-suppressed on scales larger than $m_\phi^{-1}(\phi_0)$, this implies that its effects on the large scale structure  are restricted to non-linear scales. Any cosmological observable probing linear scales, such as redshift-space distortions, should therefore see no deviation from General Relativity in these theories. 

Taken together,~(\ref{Abound}) and~(\ref{mbound}) imply that
chameleon-like scalar fields 
have a negligible effect
on density perturbations on linear scales,
and cannot account for the observed cosmic acceleration
except as some form of dark energy. 
This applies to
a broad class of chameleon, symmetron and varying-dilaton theories.
In other words,  any such model that purports to explain the observed cosmic acceleration,
and passes solar system tests, must be doing so using some form of quintessence or vacuum energy; 
the modification of gravity has nothing to do with the acceleration
phenomenon. Nonetheless, the generalized chameleon mechanism remains interesting
as a way to hide light scalars suggested by fundamental theories.
The way to test these theories is to study small scale phenomena, as summarized in the Introduction.

\subsection{No Self-Acceleration}

We first rule out self-acceleration by proving~(\ref{Abound}). Consider the equilibrium $\phi_{\rm min}$ at cosmic
mean density between redshifts $z=0$ and $z \simeq 1$, the period during which
the observed cosmic acceleration commences. Let us refer to the respective equilibrium values:
$\phi_{z=0}$ and $\phi_{z\simeq 1}$.  We are interested in theories with interesting
levels of modified gravity effects during this period; we therefore assume:
\be
\alpha(\phi) \;\gsim\; 1 \quad {\rm for} \quad 
\phi_{z\simeq 1} \le \phi \le \phi_{z=0} \,.
\label{alphaAssume}
\ee
Note that the monotonicity assumptions outlined in Sec.~\ref{reviewsec} automatically guarantees $\phi_{z\simeq 1} \le \phi_{z=0}$.
Hence $A(\phi)$ grows with time, which is a necessary condition for self-acceleration.

We are now in a position to prove~(\ref{Abound}). To do so, we must carefully examine the 
static and spherically symmetric equation of motion:
\be
\phi'' + \frac{2}{r} \phi' = V_{, \phi} + A_{, \phi} \rho\,, 
\label{staticeom}
\ee
where $'\equiv {\rm d}/{\rm d}r$. This is subject to the boundary conditions 
\be
\phi'|_{r=0} = 0\,;\qquad \phi \rightarrow_{r \rightarrow \infty} \phi_{z = 0}\,.
\ee
Although $\phi$ tends to its equilibrium value asymptotically, we make no such assumption at the origin, {\it i.e.}, 
$\phi|_{r=0} \equiv \phi_{\rm in}$ need {\it not} coincide with $\phi_{\rm min}( \rho_{\rm in})$. We distinguish 3 cases:

\begin{itemize}

\item {\it Case 1}: Suppose $V_{, \phi} + A_{, \phi} \rho \simeq 0$ at $r=0$, that is,
$\phi_{\rm in} \simeq \phi_{\rm min}( \rho_{\rm in})$. This is the thin-shell case of standard chameleons~\cite{Khoury:2003aq}.
Since $\rho_{\rm MW} \gg \rho_{z\simeq 1}$, our monotonicity assumptions imply $A(\phi_{z\simeq 1}) \ge A(\phi_{\rm in-MW})$, thus
\be
\ln \left[\frac{A(\phi_{z=0})}{A(\phi_{z\simeq 1})}\right] 
\le \ln \left[\frac{A(\phi_{z=0})}{A(\phi_{\rm in-MW})}\right]  
\lesssim 10^{-6}\,.
\label{Delphibound}
\ee
This proves (\ref{Abound}) in this case. 
 
\item {\it Case 2}: Suppose $A_{, \phi}\rho \gg -V_{, \phi}$ at $r = 0$, which is the case relevant to symmetrons~\cite{Hinterbichler:2010es}.
Given our assumption that $V_{\rm eff} = V(\phi) + A(\phi) \rho $ has a unique minimum, this implies $\phi_{\rm in}  \geq  \phi_{\rm min}( \rho_{\rm in})$.
Because $\phi'|_{r=0} = 0$, it follows from~(\ref{staticeom}) that $\phi''|_{r=0} > 0$, and thus $\phi'|_{r>0} > 0$. 
And since $\phi'$ is continuous at the surface of the object, to satisfy $\phi \rightarrow_{r \rightarrow \infty} \phi_{z = 0}$ we must therefore have
$\phi_{\rm in} < \phi_{z = 0}$. In other words, Case 2 corresponds to
\be
\phi_{\rm min}( \rho_{\rm in}) \leq \phi_{\rm in} <  \phi_{z = 0}\,.
\label{case2ineq}
\ee
Unlike Case 1, $\phi_{\rm in-MW}$ is not {\it a priori} constrained to be smaller (or greater) than 
$\phi_{z\simeq 1}$. If $\phi_{z\simeq 1} \geq \phi_{\rm in-MW}$, then as in Case 1 we are led to~(\ref{Delphibound}), and self-acceleration is ruled out.
The other possibility, $\phi_{z\simeq 1} < \phi_{\rm in-MW}$, is inconsistent with screening the Milky Way. Indeed, in this case $\phi$ falls within the range~(\ref{alphaAssume}) where $\alpha(\phi) \;\gsim\; 1$,
and~(\ref{alphadef}) can be approximated by $\nabla^2 \varphi \sim 8\pi G_{\rm N} \alpha A\rho$. Comparing with the Poisson equation $\nabla^2 \Phi_{\rm N} = 4\pi G_{\rm N} A
\rho$, it is clear the resulting scalar force is not small compared to the gravitational force, thus invalidating the screening of the Milky Way.

\item {\it Case 3}: Suppose $A_{, \phi}\rho \ll -V_{, \phi}$ at $r = 0$, that is, $\phi_{\rm in} \leq \phi_{\rm min}( \rho_{\rm in})$. In this case, all inequalities are
reversed relative to Case 2, and instead of~(\ref{case2ineq}) we conclude $\phi_{\rm min}( \rho_{\rm in}) \geq \phi_{\rm in} >  \phi_{z = 0}$. But this is inconsistent
with our assumption that $\phi_{\rm min}(\rho)$ is monotonically decreasing, hence we can ignore this case. 

\end{itemize}

To summarize, the only phenomenologically viable possibilities are Case 1, and Case 2 with $\phi_{z\simeq 1} \geq \phi_{\rm in-MW}$. In both cases we are led to~(\ref{Delphibound}).
The very small $\Delta A/A$ over cosmological time scales precludes self-acceleration.

\subsection{Bound on Compton Wavelength}

To establish the bound~(\ref{mbound}) on $m_\phi^{-1}(\phi_0)$, consider the (Einstein-frame) cosmological evolution equation:
\be
\ddot{\phi} + 3H \dot{\phi} = -V_{,\phi}  - A_{,\phi} \rho\,,
\label{phicosmo}
\ee
where $\rho$ is the total (dark matter plus baryonic) non-relativistic matter component, and $H\equiv \dot{a}_{\rm E}/a_{\rm E}$ is the Einstein-frame Hubble parameter. Since
$A(\phi)\rho\sim  H^2M_{\rm Pl}^2$ from the Friedmann equation, the density term in~(\ref{phicosmo}) exerts a significant pull on $\phi(t)$. 
The potential prevents a rapid roll-off of $\phi$ by canceling the density term to good accuracy: $V_{,\phi} \simeq - A_{,\phi} \rho$.
This cancellation must be effective over at least the last Hubble time, {\it i.e.}, $\phi$ must track adiabatically the minimum of the effective potential. 
Differentiating this relation with respect to time, and using~(\ref{mdef}) together with the conservation law $\dot{\rho} = -3H\rho$ and the Friedmann relation,
we find
\be
m^2_\phi(\phi) \simeq \frac{3HA_{,\phi}\rho}{\dot{\phi}} \sim H\frac{{\rm d} t}{{\rm d}\ln A} \alpha^2(\phi) H^2 \,.
\ee
The factor of $H^{-1}{\rm d}\ln A/{\rm d}t$ is the change of $\ln A$ over the last Hubble time, which from~(\ref{Delphibound}) is less than $10^{-6}$. Thus
\be
m^2_\phi(\phi) \;\gsim\; 10^6\alpha^2(\phi) H^2 \,.
\ee
Using~(\ref{alphaAssume}), it follows that $m_\phi^{-1}(\phi_0)\lesssim 10^{-3}H_0^{-1}\sim {\rm Mpc}$,
as we wanted to show.


\section{Chameleons and Quantum Corrections}
\label{chamquantum}

While much of the work on chameleon theories has focused on their classical description, it is crucial understand the robustness of the screening
mechanism to quantum corrections. At first sight, this question appears to be trivial --- chameleons couple to matter fields and gravitons, hence matter/graviton
loops should generate quadratically-divergent radiative corrections to the chameleon mass: $\Delta m_\phi \sim \Lambda^2/M_{\rm Pl}^2$. However, as
we will see below, the cutoff $\Lambda$ is generally so small ($\Lambda \sim {\rm meV}$) that these corrections are completely negligible. (A
similar statement holds for galileons, whose cutoff is generally even lower by a few orders of magnitude.)

\begin{figure}[tb]
\begin{center}
\includegraphics[width=6in]{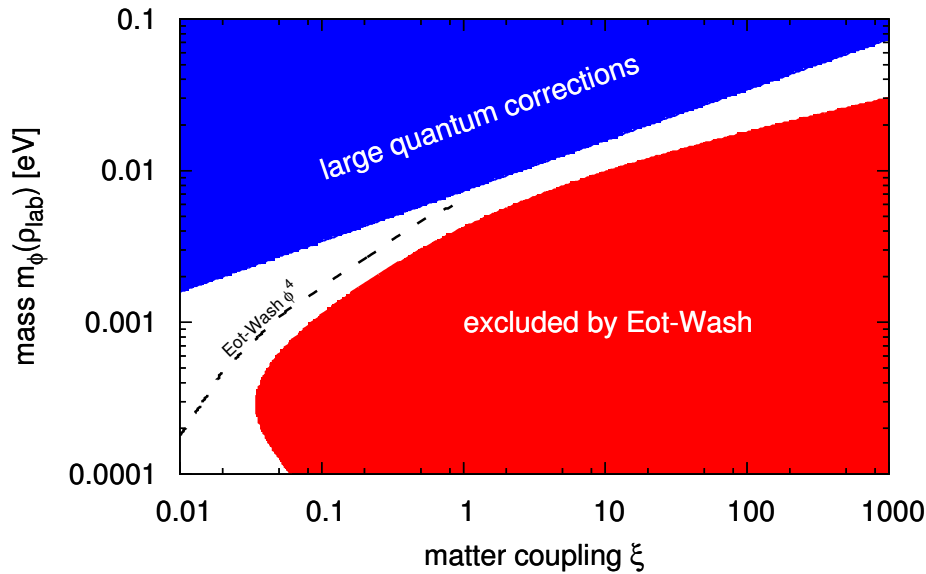}
\caption{Model-independent constraints on chameleon fields in the $\xi$,~$m_\phi$ plane with $\rholab = 10$~g/cm$^3$.  Shaded regions show loop bounds from (\ref{e:mbound}) and experimental constraints from E\"ot-Wash~\cite{Kapner:2006si}.  The dashed curve shows the direct bound on the $\phi^4$ model  for $\xi<1$~\cite{Adelberger:2006dh}, converted to $\meff$.
  \label{f:mloop_and_eotwash}}
\end{center}
\end{figure}

Below we focus on quantum corrections due to $\phi$ loops and estimate the one-loop Coleman-Weinberg correction~\cite{Upadhye:2012vh}. Since this
correction grows with increasing chameleon mass as $m^4_\phi$, it is immediately clear that quantum corrections can present problems for chameleon theories.  Chameleon screening of fifth forces operates by increasing $m_\phi$, so quantum corrections must become important above some effective mass. On the other hand, laboratory measurements place a lower bound on the effective mass. 
This causes tension between a model's classical predictivity and the predictions that it makes. Viable chameleons must tiptoe between being heavy enough to avoid fifth force constraints and remaining light enough
to keep quantum corrections under control. The results are summarized in Fig.~\ref{f:mloop_and_eotwash}.

Demanding that quantum corrections remain small compared to the classical potential, we find that the resulting ``classical'' chameleon theories cannot acquire masses larger than 
$m_\phi \sim (\xi \rho / \Mpl)^{1/3}$ at a density $\rho$ and dimensionless matter coupling $\xi$.  The upper bound on the mass, $m_\phi < 0.0073(\xi\rho/10~{\rm g}\,{\rm cm}^{-3})$~eV, is
independent of the specific form of the potential. This energy scale is interesting for dark energy models and is also accessible to upcoming fifth force experiments.  Of course, there is no requirement for Nature  to choose a model which remains a valid effective theory out to scales accessible to experiments.  However, these classical theories are the only known chameleon models with predictive power there, so our analysis can offer guidance as to the regions of the theory parameter space that future experiments should target.

For this analysis, we focus on chameleon matter couplings of the linearized form $A(\phi) = 1 + \xi \phi/M_{\rm Pl}$, where $\xi$ is constant. This linearization is valid for $|\phi| \ll  M_{\rm Pl}/\xi$, which holds for all situations of interest. The effective potential is given by
\be
V_{\rm eff}(\phi) = V(\phi) + \frac{\xi \rho(\vec x) \phi}{M_{\rm Pl}}\,.
\label{e:Veff}
\ee
Inside a sufficiently large bulk of constant matter density, the field settles to
its equilibrium value of  
\begin{equation}
V_{,\phi} (\phi_{\rm min}) =- \frac{\xi\rho}{M_{\rm Pl}} \,.
\label{eqn:phirho}
\end{equation}
As before, the potential  $V(\phi)$ so that the mass of small fluctuations, $m^2_\phi = V_{,\phi\phi}(\phi_{\rm min})$, increases with increasing matter density.

\subsection{General Bound on Chameleon Mass}

The one-loop Coleman-Weinberg correction to the classical potential $V(\phi)$, neglecting spatial variations in the field, is given as usual by
\begin{equation}
\Delta V(\phi)
=
\frac{m^4_\phi(\phi)}{64\pi^2}
\ln \left(\frac{m^2_\phi(\phi)}{\mu^2}\right)\,,
\label{e:DVloop}
\end{equation}
where $\mu$ is an arbitrary mass scale.  The one-loop corrected potential is then $\Vloop(\phi) = V(\phi) + \Delta V(\phi)$.  Even if we choose $\mu$ so that the correction 
vanishes at some fiducial density, the fact that the chameleon mass runs with field value will imply corrections at other densities. 

While large one-loop corrections do not necessarily imply the breakdown of effective field theory, in the case at hand there is no reason to expect that higher-order loop corrections will be suppressed. Thus we use the corrections arising from $\Delta V$ as a diagnostic for the breakdown of classicality. A given classical chameleon model is predictive only if these quantum corrections are small at densities of interest.

Specifically, for the chameleon mechanism to be classically predictive we require both $\Delta V_{,\phi}/V_{,\phi}$ and $\Delta V_{,\phi\phi}/V_{,\phi\phi}$ to be small across the field range of interest.
The former sets the equilibrium field value $\phi_{\rm min}$, while the latter sets the effective mass at that value. For a particular choice of $V(\phi)$, one can of course explicitly compute these quantities and
compare them with laboratory bounds. However, it is possible to cast the main physical content of the bound in a model-independent manner. Setting the log term in (\ref{e:DVloop}) to unity for simplicity,
and substituting~(\ref{eqn:phirho}), our loop criteria can be expressed as
\begin{eqnarray}
\left|\frac{\Delta V_{,\phi}}{V_{,\phi}}\right| 
&\approx& 
\left|
\frac{M_{\rm Pl}}{\xi \rho} \frac{(m_\phi^4)_{,\phi}}{64\pi^2}
\right|  
< \epsilon \,;
\nonumber
\\
\left|\frac{\Delta V_{,\phi\phi}}{V_{,\phi\phi}}\right| 
&\approx&   
\left|
\frac{(\meff^4)_{,\phi\phi}}{64\pi^2 \meff^2}
\right| 
< \epsilon \,,
\label{eqn:epsilon}
\end{eqnarray}
where $\epsilon$ should not exceed unity. 

Now, specializing~(\ref{dphimin1}) to the case of interest, we obtain
\be
\frac{{\rm d}\phi_{\rm min}}{{\rm d}\rho} = -\frac{\xi}{M_{\rm Pl}}m_\phi^{-2}(\phi_{\rm min}) \,.
\label{dphimin}
\ee
By the chain rule, it therefore follows that
\be
\frac{{\rm d}m_\phi^4}{{\rm d}\rho} = (m_\phi^4)_{,\phi} \frac{{\rm d}\phi_{\rm min}}{{\rm d}\rho} \lesssim \frac{64\pi^2\xi^2}{M_{\rm Pl}^2} \frac{\rho}{m_\phi^2}\epsilon\,,
\ee
where in the last step we have substituted~(\ref{dphimin}) and the first inequality in~(\ref{eqn:epsilon}). In other words, we have
\be
\frac{1}{\rhomat}
\frac{{\rm d} \meff^6}{{\rm d}\rhomat}  \lesssim\frac{96\pi^2 \bmat^2}{\Mpl^2}\epsilon \,.
\label{mbd1}
\ee
Similarly, the second inequality in~(\ref{eqn:epsilon}) implies
\be
\left| \frac{{\rm d}^2\meff^6}{{\rm d}\rhomat^2} \right|
\lesssim  \frac{96\pi^2 \bmat^2}{\Mpl^2}\epsilon\,.
\label{mbd2}
\ee

At laboratory density $\rholab$,~(\ref{mbd1}) and~(\ref{mbd2}) imply
\begin{equation}
\meff
\lesssim
\left(\frac{48\pi^2 \bmat^2 \rholab^2}{\Mpl^2}\right)^\frac{1}{6}
=
0.0073 \left( \frac{\bmat \rholab}{10~{\rm g\,  cm}^{-3}} \right)^\frac{1}{3}  \epsilon^\frac{1}{6} \,  {\rm eV}\,.
\label{e:mbound}
\end{equation}
For $\bmat\sim \epsilon \sim 1$ and typical densities this mass scale is close to the dark energy scale of $\rho_\Lambda^{1/4} = 0.0024$\,eV. This results from the numerical coincidence that $(\rholab/\Mpl)^{4/3} \sim \rho_\Lambda$. 
Most importantly, the dependence on $\epsilon$ is weak, and the Compton wavelength corresponding to this maximum mass, $0.027 (\bmat \rholab/10~{\rm g\, cm}^{-3})^{-1/3} \epsilon^{-1/6}$~mm, is comparable to the length scales probed by the smallest-scale torsion pendulum experiments. Given this weak dependence, henceforth we set $\epsilon = 1$, the largest value at which order-unity predictions of fifth forces can reasonably be trusted.

\begin{figure}[tb]
\begin{center}
\includegraphics[width=4in]{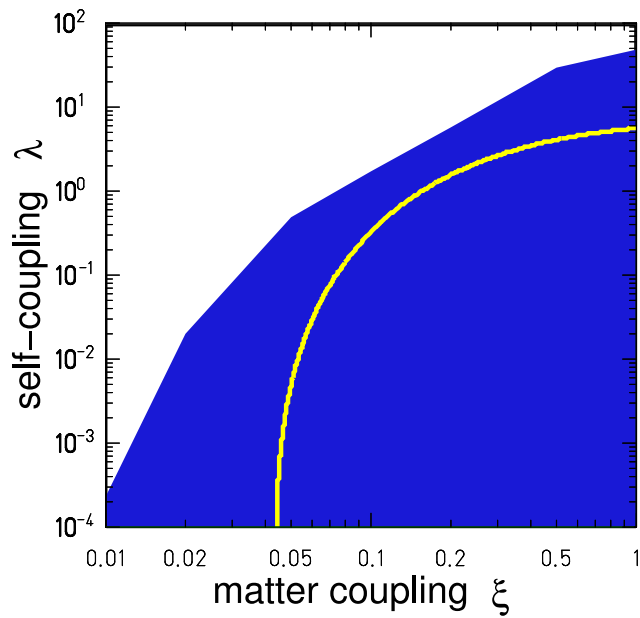}
\caption{Constraints on chameleon models with $V = \lambda\phi^4/4!$.  The shaded region shows the models excluded by~\cite{Adelberger:2006dh} while the curve shows the weaker constraints resulting from the maximum-mass approximation.}
\label{f:constraints}
\end{center}
\end{figure}

\subsection{Tension with Laboratory Constraints}

Torsion pendulum experiments such as E\"ot-Wash~\cite{Kapner:2006si} exclude fifth forces due to Yukawa scalars with constant masses $m$ over a region of the $\bmat$,~$m$ parameter space. 
If $m = \mmax$ denotes the maximum chameleon mass achieved in a given experimental set-up, then approximating the chameleon's fifth force by the Yukawa force with mass $\mmax$ should
lead to a conservative constraint on the chameleon. Indeed, since the Yukawa force has a fixed range $\mmax^{-1}$, whereas the range of the chameleon is by definition larger in parts of the experiment, 
the chameleon force should be stronger than the corresponding Yukawa force. We refer to the approximation of using $\mmax$ to
compare with data as the {\it maximum-mass} approximation. We will quantify how much bounds are improved by a direct calculation for specific potentials below.
Note that the maximum-mass approximation is used to place a minimum mass bound on $m_\phi$.

We show this  E\"ot-Wash  constraint~\cite{Kapner:2006si} on the minimum mass in Fig.~\ref{f:mloop_and_eotwash}.   We compare this to the maximum mass from the loop bound at the relevant density of  $\rholab = 10$~g/cm$^3$, working in the maximum-mass approximation. The tension between these two bounds is evident, especially near $\bmat = 1$.  A significant, but feasible, improvement in E\"ot-Wash constraints over the next several years of less than a factor of 2 in the Yukawa mass or fifth force range could eliminate all chameleon fields around $\bmat=1$ whose quantum corrections are well-controlled.
 
The maximum-mass approximation used above yields conservative but model-independent bounds on chameleon models. In the context of particular models, our approximations can be checked
against direct computation. In addition to Yukawa scalars, the E\"ot-Wash experiment has also constrained chameleon theories with $V(\phi) = \lambda\phi^4/4!$ and $\bmat<1$~\cite{Adelberger:2006dh}.
The $\phi^4$ theory is also special in that the loop bound (\ref{e:mbound}) is independent of $\rhomat$ and $\bmat$: since $\meff 
=\lambda^{1/6}(3\bmat \rhomat/\Mpl)^{1/3}$ in this case, it follows that $\lambda < 32\pi^2 \epsilon / 3 \approx 105 \epsilon$. 
In Fig.~\ref{f:mloop_and_eotwash}, we convert their constraints on $\lambda$, shown in Fig.~\ref{f:constraints}, to a bound on $\meff$.   

As expected the direct $\lambda$ bound rules out slightly more of the $\meff$ space than
our mass bound for gravitational strength $\bmat$, but there is still an allowed region which
satisfies both the loop and the laboratory bound. As also shown in Fig.~\ref{f:constraints}, the impact of our approximation on the $\lambda-\bmat$ parameter space is more pronounced since $\lambda \propto \meff^6$ but correspondingly the loop-compatible range appears larger and includes all of the space shown.   Nonetheless it is the mass  that is more closely related
to the experimental observables and even with our conservative assumptions a factor of
2 there would close the $\bmat\sim1$ window entirely in this model.

We in fact expect our constraints to be conservative for general chameleon potentials.  
To see this, consider the case of power law potentials 
\begin{equation}
V(\phi) 
=
\kappa M_\Lambda^{4-n} |\phi|^n\,,
\end{equation}
where the arbitrary mass scale $M_\Lambda$ is suggestively set to the
dark energy scale $M_\Lambda = 0.0024$\,eV, thereby making $\kappa$  a dimensionless constant.
To have a chameleon model with a bounded potential requires $n <0$ or $n>2$. 
Note that our bounds would be unchanged by adding in a constant $M_\Lambda^4$ or
a slowly varying piece to the potential that plays the role of a cosmological constant.

To model the experimental set up, consider 
a constant-density planar slab
surrounded by vacuum: $\rho(x) = \rholab$ for $x \leq 0$ and $\rho(x) = 0$ for positive $x$.
(See~\cite{Upadhye:2012qu} for a detailed calculation of chameleon profiles in torsion pendulum experiments.)
Using the exact solutions~\cite{Upadhye:2006vi,Brax:2007ak} for $V(\phi) \propto |\phi|^n$ in the vacuum $x>0$,
\begin{equation}
\phi(x)
=
\frac{\left(1-\frac{1}{n}\right)\phibulk(\rholab)}
     {\left( 1 + \sqrt{\frac{1}{2}\left| \frac{(n-2)^2(n-1)^{n-3}}{n^{n-1}} \right|} \meff(\rholab) x   \right)^\frac{2}{n-2}}\,,
\end{equation}
we can evaluate the acceleration $a_\phi = -(\bmat / \Mpl){\rm d}\phi/{\rm d}x$ of a test particle.  A Yukawa scalar $\psi$ with $m = \mmax \equiv \meff(\rholab)$ and the same matter coupling $\bmat$ will cause an acceleration $a_\psi = -(\bmat/\Mpl){\rm d}\psi/{\rm d}x$, with
\begin{equation}
\psi(x) = -\frac{\bmat \rholab}{2\mmax^2\Mpl} \exp(-\mmax x)\,.
\end{equation}
Direct comparison shows that $|a_\phi| \geq |a_\varphi|$ at $x=0$, and $|a_\phi|$ decreases more slowly than $|a_\varphi|$ for all $x \geq 0$.  Thus $|a_\phi| \geq |a_\psi|$ everywhere.  To generalize, since $\meff < \mmax$ outside the highest-density part of the experiment, the fifth force due to a chameleon falls off more slowly with distance than that due to a Yukawa scalar with $m=\mmax$.  The chameleon force is therefore larger and easier to exclude.

\begin{figure}[tb]
\begin{center}
\includegraphics[width=4in]{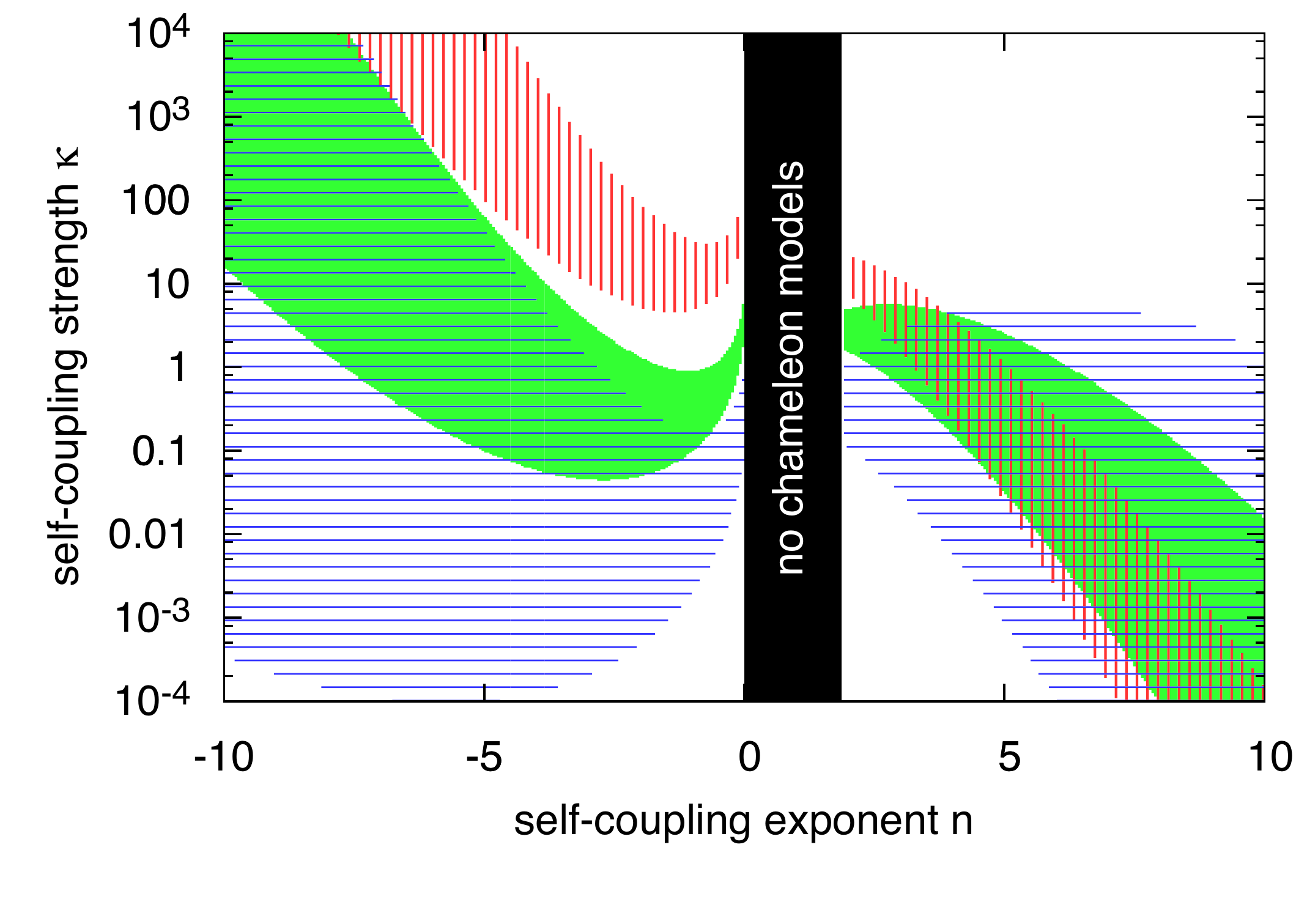}
\caption{Allowed classical chameleon models with power law potentials $V(\phi) =\kappa M_\Lambda^{4-n} |\phi|^n$, with $M_\Lambda = 0.0024$\,eV.
Blue (horizontal hatched), green (solid) and red (vertical hatched)  regions show models with $\bmat = 0.1$, $1$, and $10$, respectively, which satisfy (\ref{e:mbound}) and are consistent with~\cite{Kapner:2006si} in the maximum-mass approximation.
}
\label{f:models}
\end{center}
\end{figure}

The Yukawa mass limits can then be converted into conservative constraints
on the parameters of the power law potentials. Figure~\ref{f:models}~ shows models which are consistent with the data~\cite{Kapner:2006si} in the maximum-mass approximation and whose quantum corrections satisfy~(\ref{e:mbound}) for various $\bmat$. Although one can always find allowed models by tuning $\bmat$ to sufficiently small values, couplings of gravitational strength $\bmat \sim 1$ and higher are the most interesting for chameleon theories.  

Once again, the tension between loop corrections, which impose an upper bound on $n \kappa$, and fifth force constraints, which impose a lower bound, is evident from the figure.  A modest improvement in experimental constraints could rule out all allowed models for some range of $\bmat$. Submillimeter fifth force constraints from much denser environments, such as the $150$~g/cm${^3}$ solar center, could also rule out a substantial fraction of the models in Fig.~\ref{f:models}, though we do not have a specific probe in mind.


\section{Outlook}

Chameleons are scalar fields with remarkable properties. Through the interplay of self-interactions and coupling to matter, chameleon particles have a
mass and effective coupling strength that both depend on the ambient matter density. The manifestation of the chameleon force therefore depends sensitively
on the environment, which implies a rich phenomenology. 

In this article, we reviewed recent results that constrain the phenomenology of chameleon theories. We first reviewed two no-go theorems limiting the cosmological impact of chameleon (and symmetron)
theories. Under very general conditions, we showed that the range of the chameleon force at cosmological density today can be at most $\sim$~Mpc, and that
the conformal factor relating Einstein- and Jordan-frame scale factors is essentially constant over the last Hubble time. This implies that chameleons have negligible effect on
the linear growth of structure, and cannot account for the observed cosmic acceleration except as some form of dark energy. 

As with any no-go theorem, the key question is which of its assumptions can be circumvented? One option is to relax the assumption of adiabatic
tracking for the cosmological scalar field. While this opens up a wider range of possibilities, a generic outcome is that $\phi$ undergoes large field excursions ($\sim M_{\rm Pl}$) on cosmological time scales~\cite{Wang:2012kj}.
It is unclear whether such large field excursions are consistent with the Milky Way being screened. Another possibility is to relax 
the assumption of a single scalar field. It is likely possible to extend our no-self-acceleration theorem to a multi-field
version if $V$ and $A$ continue to be monotonic functions of $\rho$ at equilibrium. It is unclear, however, how the mass bound
would be modified in a multi-field context. Fluctuations around the effective minimum would be described by a mass matrix,
whose eigenvalues can span a wide range of scales. 

We then reviewed how keeping quantum corrections to chameleon theories under control imposes a density-dependent upper limit on the chameleon mass which is in tension with laboratory bounds on small-scale fifth forces.  This tension can be quantified in a general, model-independent way by approximating the chameleon field by a Yukawa scalar whose constant mass equals the maximum mass of the chameleon in the experiment.  Even in this conservative approximation, only a small range of viable predictive models remains for couplings around the gravitational strength, $\bmat \sim 1$, which could be excluded by a factor-of-two improvement in bounds on the range of the fifth force.

Such an improvement would test all such chameleon models, regardless of the form for their self-interaction. These models include scalar-tensor theories such as $f(R)$ theories where $\bmat = 1/\sqrt{6}$.   
Likewise they include other dark-energy motivated models where the dimensionful parameter characterizing the self-interaction is set to the dark energy scale.  

In dark-energy motivated models, the chameleon may still be invoked at lower densities, for instance to provide cosmological range forces which are sufficiently suppressed in the Solar system.  At these lower densities, the loop bound is relatively easier to satisfy; at the background matter density, for example, the range is $\meff^{-1} > 4 \times 10^5\xi^{-1/3}$m, allowing fifth forces on cosmological scales. However such models would no longer be valid effective
field theories at laboratory densities and hence would lose some of their predictive power.


\section*{Acknowledgments}
I wish to warmly thank Wayne~Hu, Lam~Hui, Amol~Upadhye and Junpu~Wang for stimulating and fruitful collaborations,
and for allowing me to borrow from our joint work in putting this review article together. I also thank Eric~Adelberger for many enlightening discussions on laboratory tests of gravity.
This work is supported in part by NSF CAREER Award PHY-1145525, NASA ATP grant NNX11AI95G, and the Alfred P. Sloan Foundation.

\end{document}